\documentclass[aps,prc,twocolumn,amsmath,superscriptaddress,floatfix,nofootinbib,showkeys]{revtex4-2}

\usepackage{amsmath,amssymb,mathtools,bm,upgreek,esint}
\usepackage{braket} 
\usepackage{graphicx}
\usepackage{adjustbox}  
\usepackage[table]{xcolor}
\usepackage{ctable}
\usepackage{setspace} 
\usepackage[normalem]{ulem}

\usepackage{verbatim} 
\usepackage{float}
\usepackage[bookmarksnumbered,bookmarksopen,colorlinks,
            linkcolor=red,
            citecolor=blue,
            urlcolor=blue,
            anchorcolor=green,
            breaklinks=true]{hyperref}

\usepackage{hyperxmp} 
\hypersetup{pdfencoding=auto}

\begin{document}
\renewcommand{\thefootnote}{\fnsymbol{footnote}}
\title{Effects of event-by-event hydrodynamic fluctuations on bottomonium dynamics in Pb--Pb collisions at $\sqrt{s_{NN}} = 5.02$ TeV}

\author{Jiamin Liu}\email{liujiamin@tju.edu.cn} 
\affiliation{Department of Physics, Tianjin University, Tianjin 300354, China}

\author{Yiyun Tang} \email{tyy@tju.edu.cn}
\affiliation{Department of Physics, Tianjin University, Tianjin 300354, China}

\author{Linyuan Wei}\email{linyuan\_wei@tju.edu.cn}
\affiliation{International Joint Institute of Tianjin University, Fuzhou,  Tianjin University, Tianjin  300072, China}

\author{Baoyi Chen}\email{baoyi.chen@tju.edu.cn}
\affiliation{Department of Physics, Tianjin University, Tianjin 300354, China}
\affiliation{International Joint Institute of Tianjin University, Fuzhou,  Tianjin University, Tianjin  300072, China}

\begin{abstract}
We investigate the effects of event-by-event hydrodynamic fluctuations on bottomonium nuclear modification factors and elliptic flow in Pb--Pb collisions at $\sqrt{s_{NN}}=5.02$ TeV. The internal evolution of the heavy quarkonium is described by a time-dependent Schr\"odinger equation with a temperature-dependent complex heavy-quark potential, while the hot QCD medium evolution is simulated using the iEBE-VISHNU event-by-event viscous hydrodynamic framework. 
By incorporating both fluctuating and smooth hot media, we find that the bottomonium nuclear modification factor $R_{\rm AA}$ is only marginally affected by event-by-event fluctuations, whereas the elliptic flow $v_2$ is systematically enhanced, with the enhancement growing from the tightly bound $\Upsilon(1{\rm S})$ to the more weakly bound $\Upsilon(2{\rm S})$ and $\Upsilon(3{\rm S})$. This enhancement arises from the more pronounced participant-plane anisotropy of the fluctuating medium relative to the smooth optical-Glauber reference geometry. These results indicate that a smooth hydrodynamic background reproduces the bottomonium $R_{\rm AA}$ but underestimates its $v_2$, so that the bottomonium $v_2$ retains a discernible imprint of event-by-event medium fluctuations.

\end{abstract}


\maketitle
\footnotetext[1]{Supported by the National Natural Science Foundation of China (NSFC) under Grant Nos.~12575149 and 12175165.}

\section{Introduction}

High-energy heavy-ion collisions at the Relativistic Heavy-Ion Collider (RHIC) and the Large Hadron Collider (LHC) create a deconfined state of strongly interacting matter, called the quark-gluon plasma (QGP), in which quarks and gluons are no longer confined inside hadrons~\cite{Bazavov:2011nk}. Heavy quarks and heavy quarkonia are among the most important probes of this medium, since they are predominantly produced in the initial hard scatterings and subsequently experience the full space-time evolution of the fireball~\cite{Matsui:1986dk,Andronic:2015wma,Rothkopf:2019ipj,Zhao:2020jqu,He:2022ywp}. In particular, bottomonium states provide a clean hard probe because the bottom-quark mass is much larger than the typical temperatures of the QGP, which makes a nonrelativistic description for heavy quarkonium well justified.

From the theoretical side, the evolution of heavy quarkonium in hot QCD matter can be formulated in terms of an in-medium heavy-quark potential within effective field theory approaches such as potential nonrelativistic QCD~\cite{Brambilla:1999xf,Brambilla:2016wgg}. The corresponding potential becomes complex in the deconfined medium~\cite{Margotta:2011ta,Brambilla:2020qwo,Wen:2022yjx}. Its real part encodes color screening and modifies the binding structure of quarkonium states, while its imaginary part describes in-medium dissociation induced by interactions with the surrounding light partons. Combined with the time-dependent Schr\"odinger equation, this framework provides a dynamical and microscopically motivated description of bottomonium suppression in nuclear collisions.

At the same time, relativistic hydrodynamics has achieved remarkable success in describing the soft sector of heavy-ion collisions, indicating that the QGP behaves as an almost perfect fluid with very small specific shear viscosity~\cite{Schenke:2010nt,Song:2010mg,Heinz:2013th}. It is now well established that the medium created in each collision event is not smooth. Instead, event-by-event fluctuations in the initial entropy deposition generate irregular spatial structures and local hot spots, which are then converted by the hydrodynamic expansion into the observed anisotropic flow of final-state hadrons~\cite{Pang:2012he,Alver:2010gr,Schenke:2012wb,Gardim:2011xv,Qiu:2011iv}. For light hadrons, such fluctuations are essential for understanding higher-order flow harmonics and event-plane correlations.

For heavy quarkonia, however, the quantitative impact of realistic hydrodynamic fluctuations is still less explored. Early studies already showed that initial-state fluctuations can noticeably affect the suppression pattern of excited bottomonium states, even when the effect on the ground state remains relatively small~\cite{Song:2013tla}. More recently, bottomonium suppression and elliptic flow in fluctuating hydrodynamic backgrounds have also been investigated within real-time quantum-evolution frameworks, indicating that the overall influence of fluctuating initial conditions can depend on both the modeling of the medium and the treatment of quarkonium dynamics~\cite{Islam:2020bnp,Alalawi:2022gul}. These developments make it timely to perform a systematic comparison between bottomonium dynamics in the event-by-event fluctuating and smooth hydrodynamic backgrounds. In contrast to approaches based on a sharp dissociation temperature or phenomenological inelastic cross sections, the present calculation describes bottomonium suppression through the direct time evolution of the wave function with a complex in-medium heavy-quark potential. In this framework, local temperature variations modify the accumulated real-time evolution and damping
of the bottomonium wave packet along its trajectory, rather than removing a state instantaneously once the local temperature exceeds a prescribed threshold. The purpose of the present work is to quantify whether a smooth hydrodynamic background can reproduce bottomonium \(R_{AA}\) and \(v_2\) obtained from event-by-event fluctuating hydrodynamics within the same time-dependent Schr\"odinger-equation framework and the same complex-potential uncertainty band. We find that the smooth background reproduces $R_{\rm AA}$ within the theoretical uncertainty, while it systematically underestimates $v_2$.

Compared with previous real-time quantum-evolution studies of bottomonium in fluctuating hydrodynamic backgrounds, the present work focuses on a direct comparison between event-by-event fluctuating hydrodynamic media and a smooth optical-Glauber reference background within a Schr\"odinger-equation framework using a phenomenologically constrained complex heavy-quark  potential. We analyze both $R_{AA}$ and $v_2$ for $\Upsilon(1S)$, $\Upsilon(2S)$, and $\Upsilon(3S)$ in different centralities. The remainder of this paper is organized as follows. In Sec.~II, we introduce the Schr\"odinger framework and the in-medium heavy-quark potential. In Sec.~III, we describe the event-by-event hydrodynamic background and its validation against soft-hadron data. In Sec.~IV, we present the numerical results for bottomonium suppression and flow and discuss their physical implications.

\section{Potential model for bottomonium}

Given the large heavy-quark mass, the internal evolution of bottomonium can be described by the time-dependent Schr\"odinger equation. The radial component of the heavy quark dipole wave function can be separated as follows:
\begin{align}
i\hbar \frac{\partial}{\partial t} \psi(r, t) = 
\left( -\frac{\hbar^2}{2m_\mu} \frac{\partial^2}{\partial r^2} + V(T,r) + \frac{l(l+ 1)\hbar^2}{2m_\mu r^2} \right) \psi(r, t),
\label{eq:sch}
\end{align}
where $m_{\mu}=m_b/2$ is the reduced mass and $m_b=4.62~\mathrm{GeV}$ is the bottom-quark mass. The reduced radial wave function is defined by $\psi(r,t)=rR(r,t)$, with $R(r,t)$ the radial wave function of the quarkonium state. The orbital angular momentum quantum number is denoted by $l$, with $l=0$ and $l=1$ corresponding to $S$-wave and $P$-wave channels, respectively. The time dependence in Eq.~(\ref{eq:sch}) enters through the local medium temperature $T$, which evolves along the trajectory of the propagating $b\bar b$ dipole in the expanding QGP background.

The in-medium heavy-quark potential is taken to be complex, $V(T,r)=V_R(T,r)+V_I(T,r)$. The real part $V_R$ describes the screened interaction between the heavy quark and antiquark, while the imaginary part $V_I$ accounts for in-medium dissociation. For the real part, we use a screened Cornell-type parametrization,
\begin{align}
V_R(T,r)
&=
-\alpha\left(\mu_D+\frac{e^{-\mu_D r}}{r}\right)
+\frac{2\sigma}{\mu_D}
-\frac{\sigma e^{-\mu_D r}(2+\mu_D r)}{\mu_D},
\label{eq:VR}
\end{align}
where the Coulomb coupling and string tension are taken to be $\alpha=\pi/12$ and $\sigma=0.2\,\mathrm{GeV}^2$, respectively~\cite{Satz:2005hx}. $\mu_D$ is the Debye screening mass. The Debye mass is parametrized~\cite{Liu:2026uav} as $\mu_D=a_4 T\sqrt{4\pi N_c\left(1+\frac{N_f}{6}\right)\frac{\alpha}{3}}$, with $N_c=3$ and $N_f=3$. The parameter $a_4$ controls the overall magnitude of color screening in the medium. For the imaginary part, we adopt the parametrization~\cite{Liu:2026uav}
\begin{equation}
V_I
=
- i T^{a_0}\left(a_1 {\bar r}+a_2 {\bar r}^{a_3}\right),
\label{eq:VI}
\end{equation}
where \(\bar r=r/{\rm fm}\) is the dimensionless radial distance. \(a_0\) controls the temperature dependence, while \(a_1\), \(a_2\), and \(a_3\) determine the radial dependence of the thermal width. The real part controls the in-medium binding structure, while the imaginary part induces norm loss of the color-singlet wave packet. In the present work, we explicitly propagate only the color-singlet component of the \(b\bar b\) pair. Color-octet or unbound configurations are not treated as independent degrees of freedom, but are effectively included through the absorptive imaginary part of the potential as probability loss from the singlet sector. The lost components do not contribute to the final projection onto vacuum bottomonium eigenstates. The inverse octet-to-singlet transition, or regeneration contribution, is not included in the present calculation. The in-medium heavy-quark potential is used in the Schr\"odinger equation during the QGP phase, defined by \(T>T_c\), with \(T_c=0.17\) GeV. 

\begin{figure}[htbp]
\centering
\includegraphics[width=0.48\linewidth]{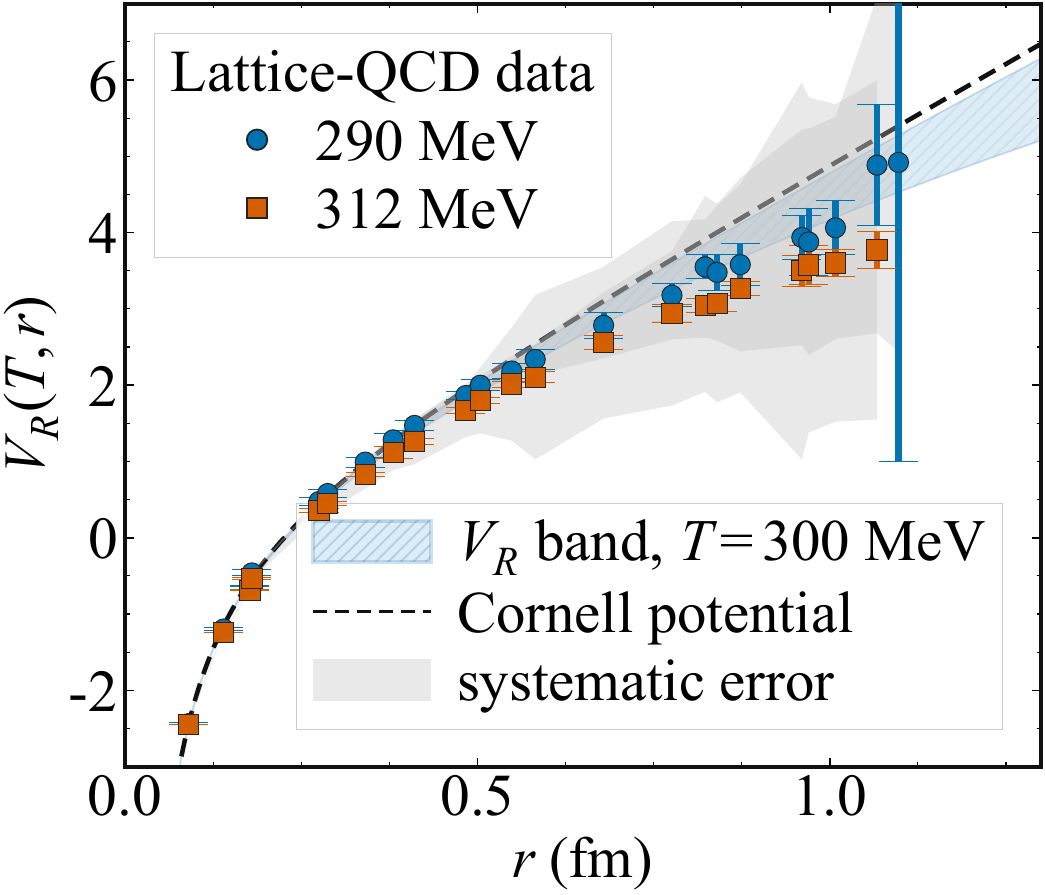}
\includegraphics[width=0.48\linewidth]{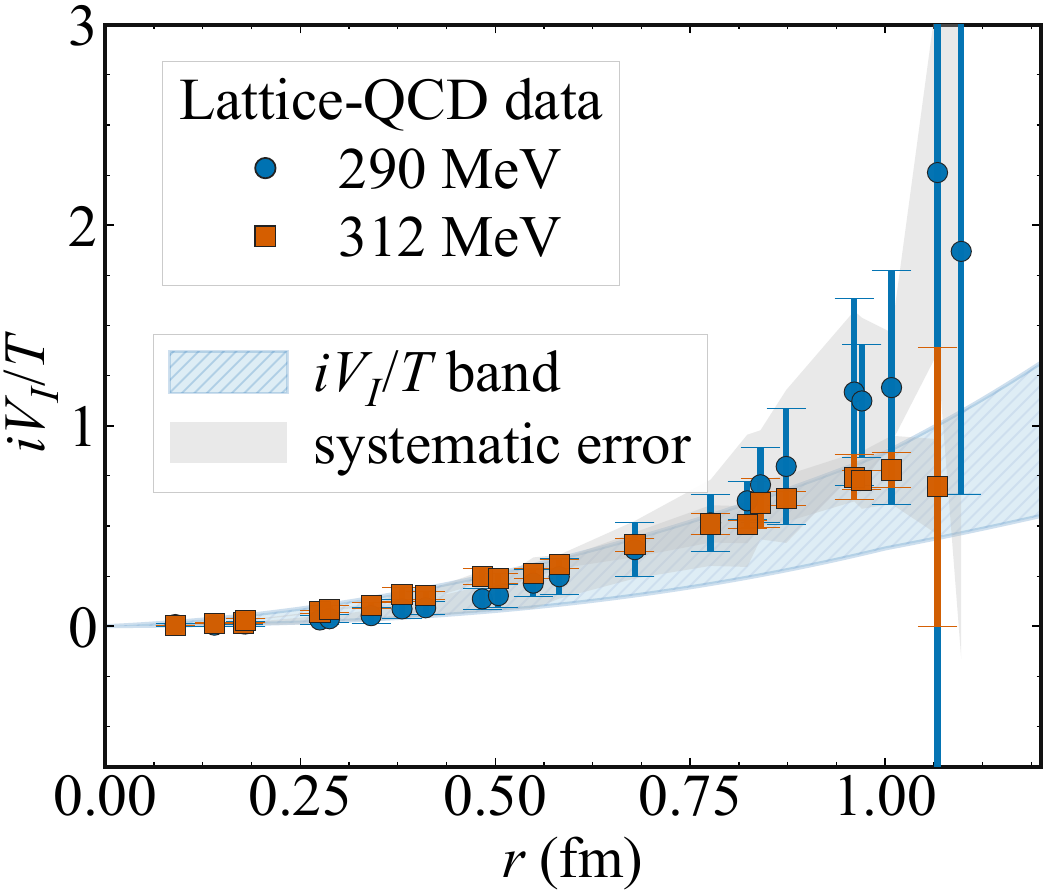}
\caption{In-medium heavy-quark potential used in the Schr\"odinger evolution. 
The left panel shows the screened real part $V_R(T,r)$ together with the vacuum Cornell potential, while the right panel shows the scaled imaginary part $iV_I/T$.
The temperature of the bands in $V_R$ and $V_I$ is taken as $T=300$ MeV.  The lattice-QCD data are plotted as a comparison~\cite{Burnier_2017}. 
}
\label{fig:Potential}
\end{figure}

In the hadronic gas phase ($T < T_c$), we neglect medium effects and adopt the vacuum Cornell potential in the Schr\"odinger equation. The parameter ranges of the complex potential are taken from the Bayesian extraction in Ref.~\cite{Liu:2026uav}, namely
$a_0\in[1.001,\,1.407]$, $a_1\in[0.034,\,0.180]$, $a_2\in[0.354,\,0.687]$, $a_3\in[2.000,\,2.579]$, and $a_4\in[0.100,\,0.319]$. These result in an uncertainty in the $V_R$ and $V_I$ plotted in Fig.~\ref{fig:Potential}, which will be used in the following bottomonium calculations. The real component of the heavy quark potential closely resembles the vacuum Cornell potential, aligning well with findings from both Bayesian inference~\cite{b8n3-yrq1} and deep learning approaches~\cite{Shi:2021qri,Liu:2026uav}. The suppression of quarkonium states is primarily driven by the imaginary part of the potential. 

In relativistic heavy-ion collisions, heavy quarks are produced predominantly in initial hard scatterings. Accordingly, the initial spatial density of $b\bar b$ dipoles is taken to be proportional to the density of binary nucleon-nucleon collisions~\cite{Miller:2007ri},
\begin{equation}
\rho_{b\bar b}(\mathbf{x}_T)
\propto
T_A(\mathbf{x}_T-\mathbf{b}/2)\,T_B(\mathbf{x}_T+\mathbf{b}/2),
\end{equation}
where $T_A$ and $T_B$ are the thickness functions of the two colliding Pb nuclei and $\mathbf{b}$ is the impact parameter.
In the present calculation, this optical-Glauber binary-collision profile is used to sample the initial \(b\bar b\) positions in both the fluctuating and smooth medium calculations. Thus, the hard-production baseline is kept the same in the two cases, and the comparison focuses on the effect of event-by-event fluctuations of the QGP temperature field. Event-wise fluctuations of the hard-production profile and their correlation with the MC-Glauber medium fluctuations are not included. Their total momentum is assumed to remain unchanged during the in-medium evolution, so that the medium affects only the internal wave function. The transverse momentum of the initially produced $b\bar b$ dipoles is sampled from a power-law distribution motivated by the measured bottomonium spectra in $pp$ collisions~\cite{b8n3-yrq1},
\begin{equation}
\frac{d N}{2 \pi p_T dp_T}
=
\frac{n-1}{\pi (n-2)\langle p_T^2\rangle}
\left(
1+\frac{p_T^2}{(n-2)\langle p_T^2\rangle}
\right)^{-n} ,
\end{equation}
where $\langle p_T^2\rangle=80~(\mathrm{GeV}/c)^2$ and $n=2.5$.

The time-dependent Schr\"odinger equation is solved on a radial grid using an implicit finite-difference scheme. At each time step, the discretized radial equation leads to a tridiagonal linear system, which is solved with the tridiagonal matrix algorithm~\cite{Liu:2026uav}. By employing the method described above, we evolve each $b\bar{b}$ dipole event-by-event along unique trajectories, initializing them with random initial positions and total momenta. The Schr\"odinger evolution continues until the dipole exits the QGP medium at time $t_f$. At this stage, the survival fraction of each bottomonium eigenstate $(n,l)$ is calculated as $|c_{nl}(t_f)|^2$, where the expansion coefficients are obtained by projecting the final reduced radial wave function onto the corresponding vacuum eigenstate, $c_{nl}(t_f)=\langle \phi_{nl}|\psi(t_f)\rangle .$ Here, $\phi_{nl}(r)$ denotes the radial wave function of the vacuum eigenstate with principal quantum number $n$ and orbital angular momentum quantum number $l$. The prompt bottomonium yield is obtained by including feed-down contributions from higher excited states~\cite{Islam:2020bnp,b8n3-yrq1}. In the present calculation, we evolve the coupled set $\Upsilon(1S)$, $\chi_b(1P)$, $\Upsilon(2S)$, $\chi_b(2P)$, and $\Upsilon(3S)$, where the $\chi_b(nP)$ states denote the spin-averaged triplets. The prompt nuclear modification factor of a final observed state $i$ is calculated as
\begin{equation}
R_{AA}^{\rm prompt}(i)
=
\frac{
\sum_j {\cal B}_{j\to i}\,
\sigma_{\rm direct}(j)\,
R_{AA}^{\rm direct}(j)
}{
\sum_j {\cal B}_{j\to i}\,
\sigma_{\rm direct}(j)
},
\end{equation}
where \({\cal B}_{j\to i}\) is the inclusive feed-down branching fraction
from state \(j\) to the observed final state \(i\). The feed-down branching
fractions are taken from Ref.~\cite{Islam:2020bnp}, and the direct production cross sections listed in Table~\ref{pp-cross-section} are obtained by inverting the corresponding feed-down relations.

\begin{table}[t]
\centering
\caption{Prompt and direct bottomonium production cross sections in the central rapidity region of $\sqrt{s}=5.02~{\rm TeV}$ $pp$ collisions~\cite{CMS:2018zza,Lansberg:2019adr,Brambilla:2020qwo,CMS:2013qur}. }
\label{pp-cross-section}
\begin{tabular}{lccccc}
\hline
State 
& $\Upsilon(1S)$ 
& $\chi_b(1P)$ 
& $\Upsilon(2S)$ 
& $\chi_b(2P)$ 
& $\Upsilon(3S)$ \\
\hline
$\sigma_{\rm exp}$ (nb)    
& 57.6 & 33.51 & 19.0 & 29.42 & 6.8 \\
$\sigma_{\rm direct}$ (nb) 
& 37.97 & 44.20 & 18.27 & 37.68 & 8.21 \\
\hline
\end{tabular}
\end{table}

The nuclear modification factor is calculated from the event-averaged prompt yield as
\begin{equation}
R_{AA}(p_T)=
\frac{
\left\langle
\int d\varphi\,
\frac{dN^{AA}_{e}}{dp_T d\varphi}
\right\rangle_{\rm ev}
}{
\langle N_{\rm coll}\rangle
\int d\varphi\,
\frac{dN^{pp}}{dp_T d\varphi}
}.
\label{eq:raa-def}
\end{equation}
Here $dN^{AA}_{e}/dp_T d\varphi$ denotes the prompt bottomonium yield obtained in a single hydrodynamic event, and $\langle\cdots\rangle_{\rm ev}$ denotes the average over fluctuating hydrodynamic events. For the smooth reference background, the same expression is used without the event average. The quantity $dN^{pp}/dp_T d\varphi$ is the corresponding prompt yield in $pp$ collisions, and $\langle N_{\rm coll}\rangle$ denotes the number of binary nucleon-nucleon collisions in the corresponding centrality class.

The elliptic flow coefficient is calculated with respect to the second-order event-plane angle as
\begin{equation}
v_2(p_T)
=
\frac{
\left\langle
\int d\phi\,
\cos\left[2\left(\phi-\Psi_2^{(e)}\right)\right]\,
\frac{dN_{AA}^{e}}{dp_T d\phi}
\right\rangle_{\rm ev}
}{
\left\langle
\int d\phi\,
\frac{dN_{AA}^{e}}{dp_T d\phi}
\right\rangle_{\rm ev}
}.
\label{eq:v2_def}
\end{equation}
Here \(\phi\) denotes the azimuthal angle of the bottomonium transverse
momentum, and \(\Psi_2^{(e)}\) is the second-order event-plane angle of the
\(e\)-th hydrodynamic event. After rotating each event to its own event-plane frame, \(\phi'=\phi-\Psi_2^{(e)}\), the angular factor can be equivalently written as $\cos(2\phi')=\frac{p_x^{\prime 2}-p_y^{\prime 2}}{p_x^{\prime 2}+p_y^{\prime 2}}$, where \(p_x'\) and \(p_y'\) are the transverse-momentum components in the event-plane-aligned frame. The elliptic flow coefficient is expected to reflect the anisotropy of the medium distribution arising from event-by-event fluctuations, which is encoded through the interactions between heavy quarkonia and the hot deconfined medium.

\section{Fluctuating hot medium}

To simulate the space-time evolution of the hot QCD medium, we employ the iEBE-VISHNU framework, which provides event-by-event viscous hydrodynamic simulations for relativistic heavy-ion collisions~\cite{Shen:2014vra}. In the present work, the fluctuating initial conditions are generated by the superMC module based on the Monte Carlo Glauber model. The initial entropy density is taken as a linear combination of participant and binary-collision contributions,
\begin{equation}
s(x,y)
=
s_0\left[
(1-\alpha)\frac{N_{\rm part}(x,y)}{2}
+\alpha N_{\rm coll}(x,y)
\right],
\end{equation}
where $N_{\rm part}(x,y)$ and $N_{\rm coll}(x,y)$ are the local densities of participant nucleons and binary collisions, respectively. The parameter $\alpha$ controls the relative admixture of the two components, and the normalization factor $s_0$ is adjusted to reproduce the charged-particle multiplicity given by experiments.

Event-by-event fluctuations are implemented through the stochastic sampling of nucleon positions together with additional entropy deposition fluctuations for each participant, modeled by a Gamma distribution in the superMC framework~\cite{Shen:2014vra}. The Gamma fluctuation parameter is taken to be $0.75$, and the two-component mixing parameter is set to $\alpha=0.118$~\cite{ALICE-exp}.

The subsequent evolution of the medium is described by a $(2+1)$-dimensional longitudinally boost-invariant viscous hydrodynamic framework in the Denicol--Niemi--Moln\'ar--Rischke formulation~\cite{Denicol:2012cn}, as implemented in the iEBE-VISHNU package~\cite{Shen:2014vra}. We use a constant specific shear viscosity $\eta/s=0.08$, neglect bulk viscosity, start the hydrodynamic evolution at $\tau_0=0.6~\mathrm{fm}/c$, and adopt the s95p-PCE equation of state~\cite{Huovinen:2009yb}. The hydrodynamic evolution is terminated at the decoupling energy density $e_{\rm dec}=0.18~\mathrm{GeV/fm^3}$. 
Particle emission at the decoupling surface is implemented through the Cooper--Frye prescription with shear viscous corrections to the distribution function~\cite{Cooper:1974mv}; the subsequent UrQMD hadronic cascade is not included.

In $\sqrt{s_{NN}}=5.02~\mathrm{TeV}$ Pb--Pb collisions, event-by-event fluctuating backgrounds are generated in several centrality classes and then combined into the centrality intervals used in the bottomonium analysis. In the present study, the bottomonium observables are primarily presented for the 0--20\% and 30--50\% centrality intervals.

\begin{figure*}[t]
\centering
\includegraphics[width=0.9\linewidth]{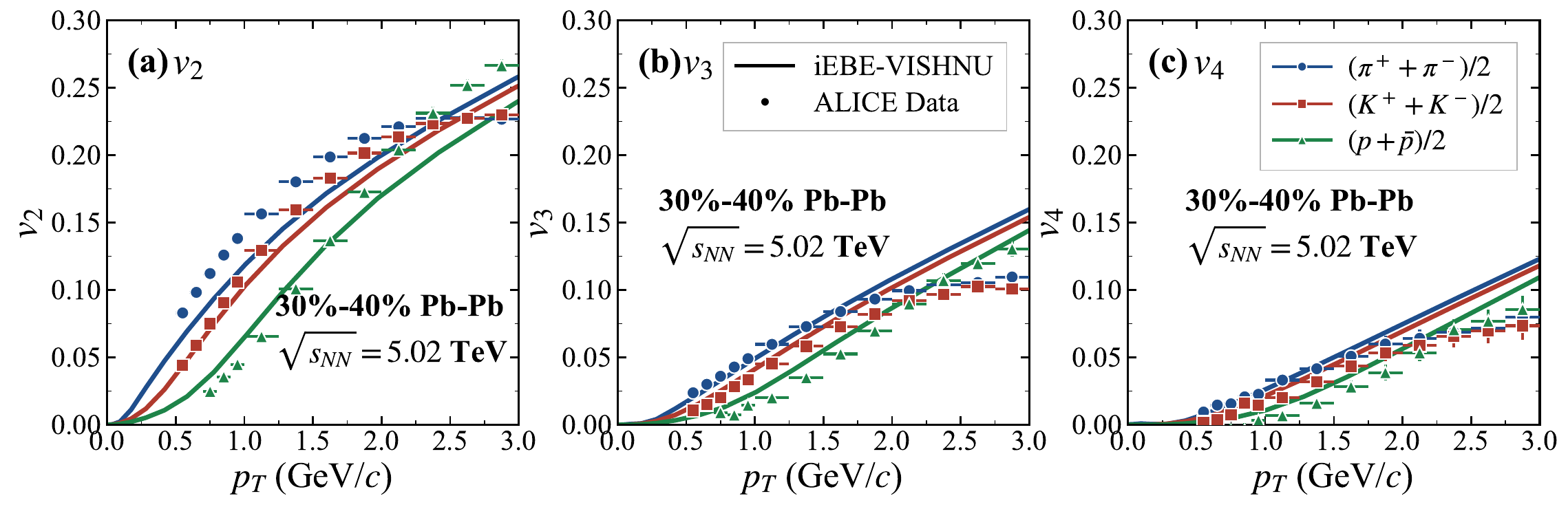}
\caption{Elliptic flow coefficient $v_2$ and higher-order flow coefficients $v_3$ and $v_4$ of identified light hadrons, $(\pi^+ + \pi^-)/2$, $(K^+ + K^-)/2$, and $(p+\bar{p})/2$, in 30--40\% Pb--Pb collisions at $\sqrt{s_{NN}}=5.02~\mathrm{TeV}$. The solid lines represent the iEBE-VISHNU calculations, while the symbols denote the ALICE data~\cite{ALICE:2018yph}.}
\label{fig:v2_light}
\end{figure*}

Before applying the hydrodynamic background to bottomonium evolution, we briefly check that the initial setup in the previous sections provides a reasonable description of representative soft-hadron observables. The final multiplicities of light hadrons from the hydrodynamic model and the experiments are listed in Table \ref{tab:multiplicity}.

\begin{table}[htbp]
\centering
\caption{Mean charged-hadron multiplicity in Pb--Pb collisions at $\sqrt{s_{NN}}=5.02~\mathrm{TeV}$, compared with the ALICE measurements~\cite{ALICE-exp}.}
\label{tab:multiplicity}
\begin{tabular}{ccc}
\hline
Centrality & $\langle d N_{ch}/d\eta \rangle_{\rm th}$ & $\langle d N_{ch}/d\eta \rangle_{\rm exp}$ \\
\hline
0--5\%   & 1996.4  & $1943\pm 56$ \\
30--40\% & 505.6   & $512\pm 15$ \\
50--60\% & 172.1  & $183\pm 8$ \\
80--90\% & 16.5    & $17.5\pm 1.8$ \\
\hline
\end{tabular}
\end{table}

The flow coefficients \(v_2\), \(v_3\), and \(v_4\) of identified light hadrons obtained after Cooper--Frye particlization are shown in Fig.~\ref{fig:v2_light}. In the present calculation, no subsequent UrQMD hadronic cascade is included. Therefore, Fig.~\ref{fig:v2_light} should be viewed as a consistency check of the hydrodynamic background rather than as a precision soft-sector calibration. The charged-particle multiplicities and the overall magnitudes of the representative soft-hadron flow coefficients are reasonably described, which is sufficient for the purpose of the present study, namely comparing bottomonium evolution in fluctuating and smooth QGP backgrounds within the same hydrodynamic setup.

\begin{figure}[t]
\centering
\includegraphics[width=0.48\linewidth]{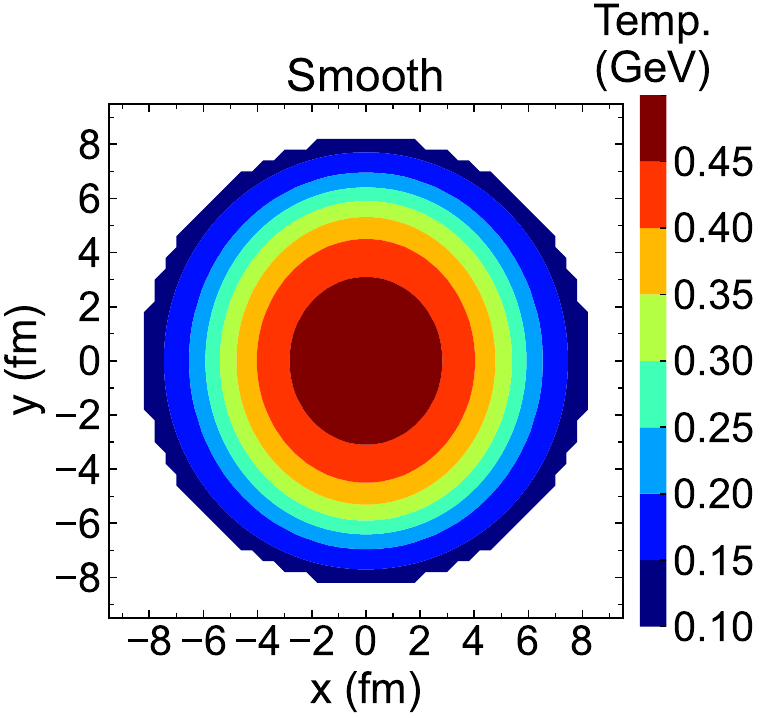}
\includegraphics[width=0.48\linewidth]{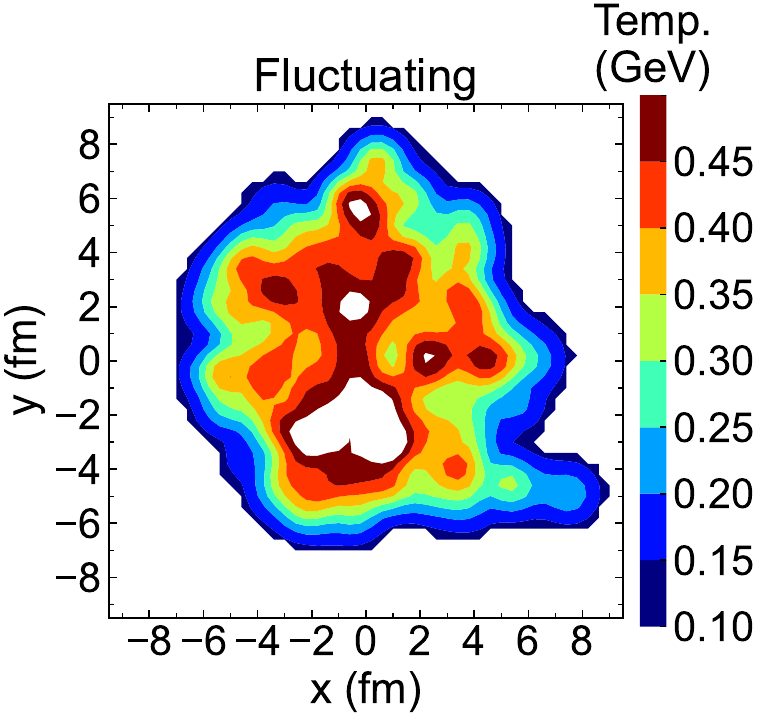}
\caption{Representative temperature contour maps at the time \(\tau=0.6~\mathrm{fm}/c\) for smooth and one fluctuating hot medium in the 0--20\% centrality class of Pb--Pb collisions at $\sqrt{s_{NN}}=5.02~\mathrm{TeV}$.}
\label{fig:hydro_maps}
\end{figure}
To disentangle the effect of event-by-event fluctuations from that of the averaged medium evolution, we compare bottomonium observables obtained in two different types of hydrodynamic backgrounds: a fluctuating event-by-event background and a smooth reference background. In the former case, each bottomonium trajectory is evolved in an individual hydrodynamic event with its own local temperature inhomogeneities. In the latter case, the evolution is performed in a smooth background corresponding to the same centrality interval. For the smooth reference calculation, the initial entropy-density profile is constructed from an optical-Glauber distribution within the same centrality class. The normalization of the smooth initial condition is chosen to reproduce the same mean final charged-particle multiplicity as in the event-by-event fluctuating hydrodynamic calculation. The subsequent hydrodynamic evolution is performed with the same transport coefficients, equation of state, starting time, and decoupling condition as in the fluctuating calculation.

In a fluctuating hydrodynamic background, the second-order event-plane angle is determined from its initial entropy density according to the standard participant-plane convention used in event-by-event hydrodynamic calculations~\cite{Alver:2010gr,Qiu:2011iv},
\begin{equation}
\Psi_2
=
\frac{1}{2}
{\rm atan2}
\left(
-\langle r^2\sin 2\varphi_s\rangle_s,
-\langle r^2\cos 2\varphi_s\rangle_s
\right),
\end{equation}
where \(\varphi_s\) denotes the spatial azimuthal angle and
\(\langle\cdots\rangle_s\) represents an entropy-density-weighted average in the transverse plane. The $\Psi_2$ in each fluctuating event will be calculated and used in the formula of bottomonium $v_2$.

Figure~\ref{fig:hydro_maps} shows a representative comparison between a smooth hydrodynamic background and an event-by-event fluctuating background in the 0--20\% centrality class. 
Compared to the smooth reference profile, the fluctuating event contains both localized hot spots and relatively colder regions. Therefore, the effect of event-by-event fluctuations on bottomonium survival is determined by the competition between enhanced dissociation in hotter regions and reduced dissociation in colder regions along the propagation path.

\section{Bottomonium suppression and flow in fluctuating media}

In this section, we present the effects of event-by-event hydrodynamic fluctuations on bottomonium suppression and elliptic flow in Pb--Pb collisions at $\sqrt{s_{NN}}=5.02~\mathrm{TeV}$. We compare calculations based on fluctuating and smooth hydrodynamic backgrounds for $\Upsilon(1S)$, $\Upsilon(2S)$, and $\Upsilon(3S)$ in the 0--20\% and 30--50\% centrality intervals. The comparison allows us to examine how the final observables depend simultaneously on the centrality class, the binding strength of the bottomonium state, and the competition between local temperature inhomogeneities and the global geometric anisotropy of the medium. In all result figures, the shaded bands represent the uncertainty propagated from the allowed parameter ranges of the complex heavy-quark potential.

\begin{figure*}[htbp]
\centering
\includegraphics[width=0.3\linewidth]{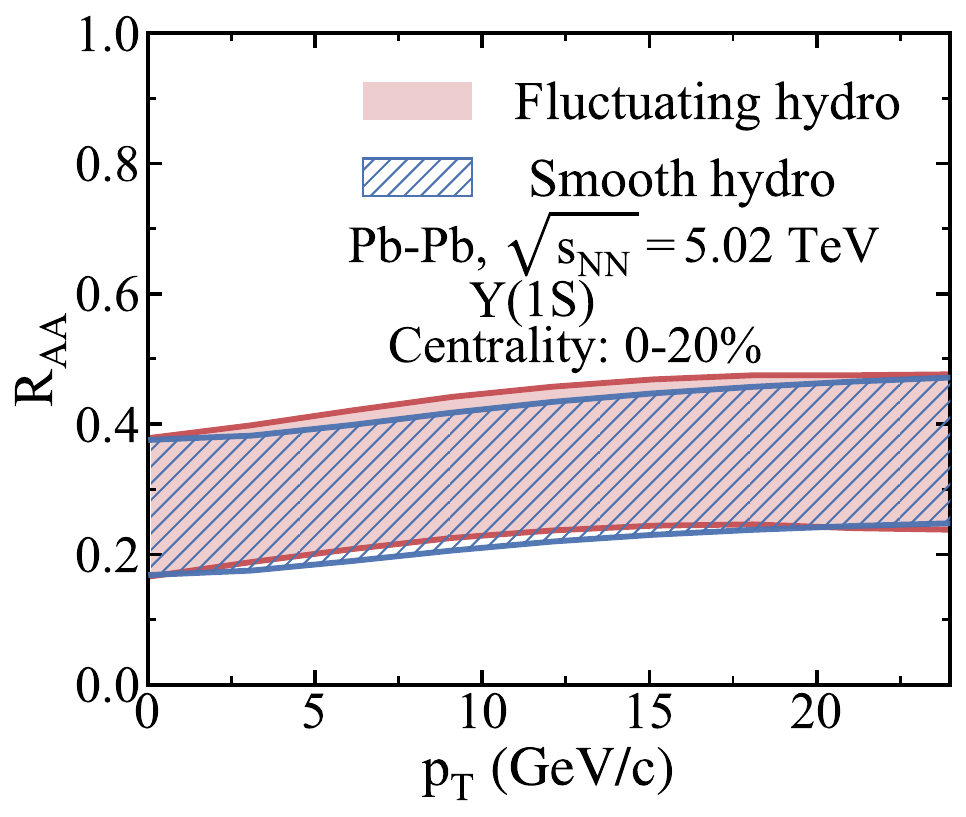}
\includegraphics[width=0.3\linewidth]{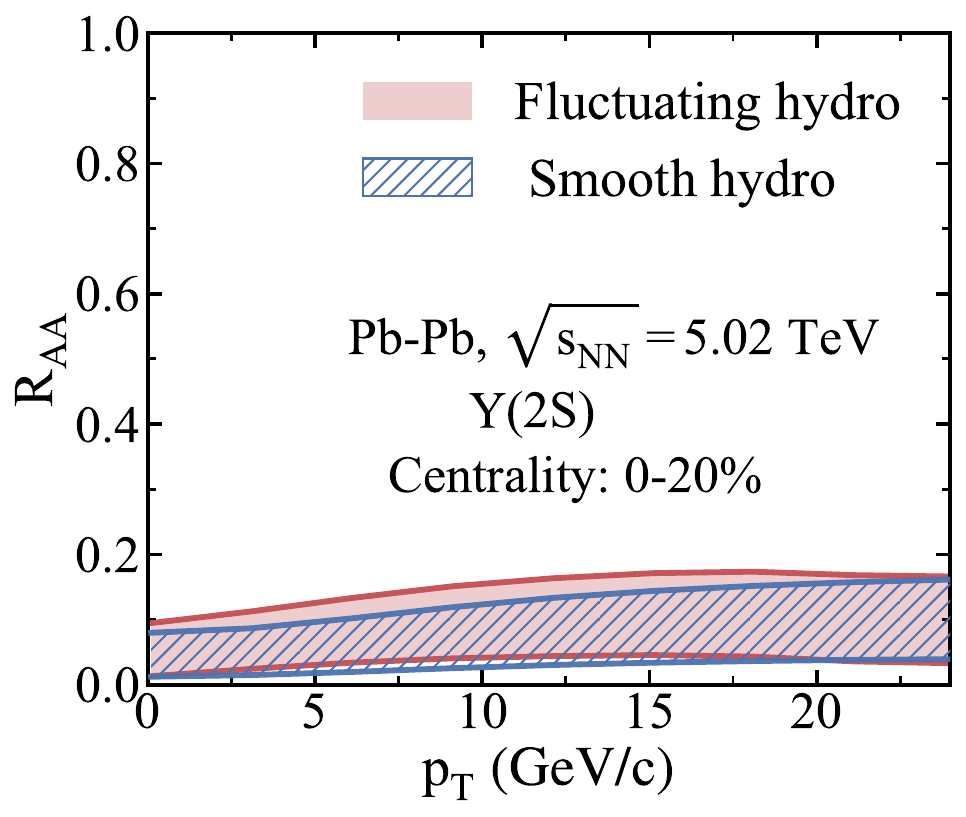}
\includegraphics[width=0.3\linewidth]{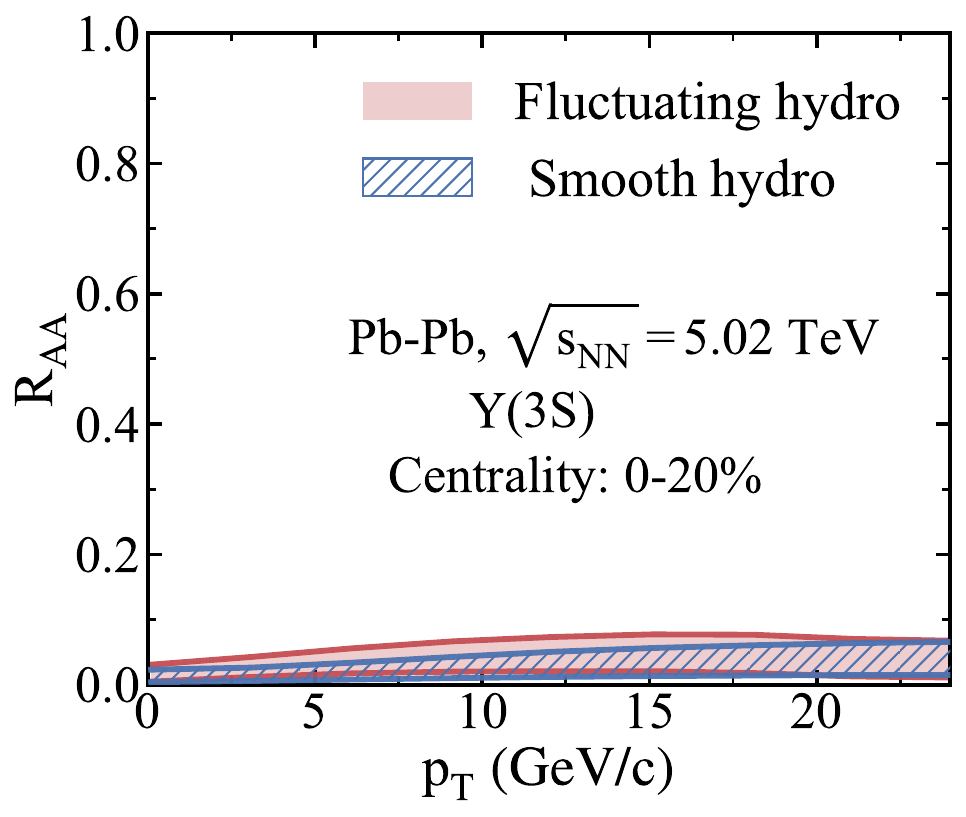}
\includegraphics[width=0.3\linewidth]{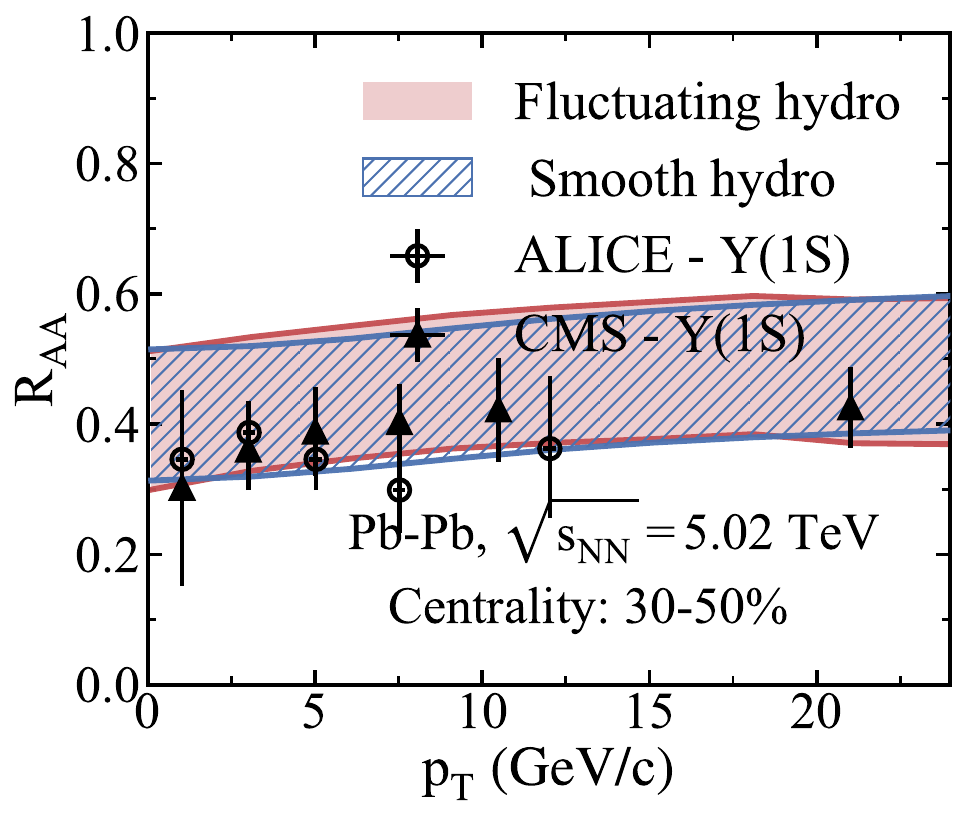}
\includegraphics[width=0.3\linewidth]{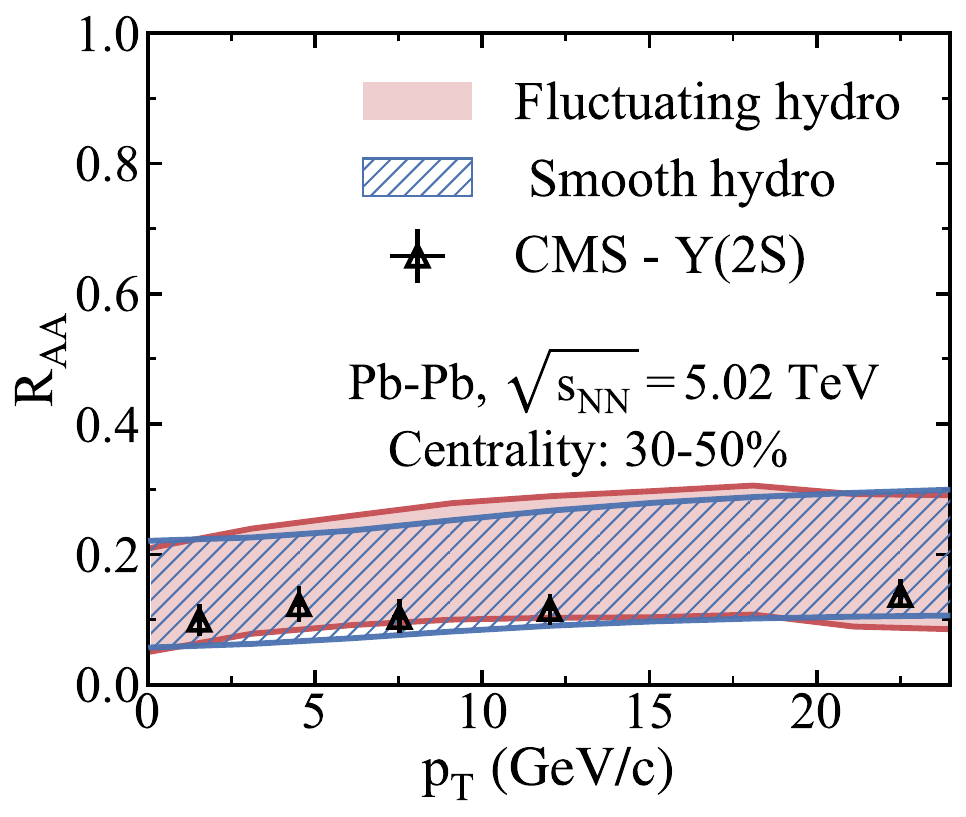}
\includegraphics[width=0.3\linewidth]{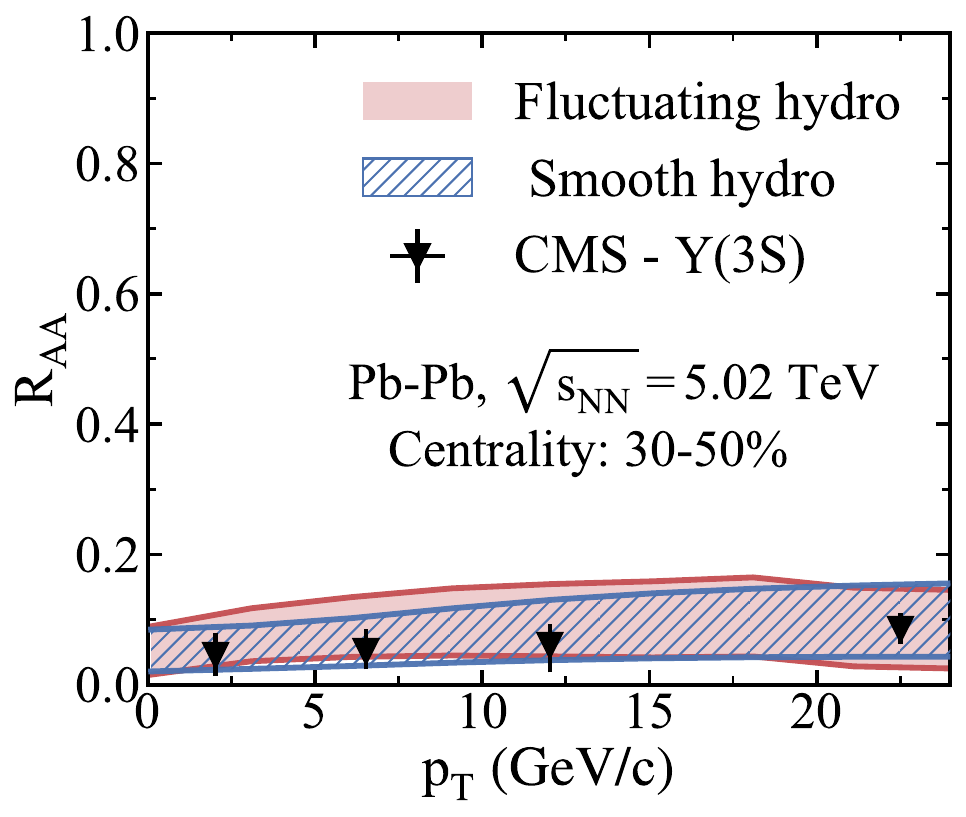}

\caption{Comparison of bottomonium $R_{AA}$ in fluctuating and smooth hydrodynamic backgrounds for       $\Upsilon(1S)$, $\Upsilon(2S)$, and $\Upsilon(3S)$ in the 0--20\% (upper panels) and 30--50\% (lower panels) centrality intervals of Pb--Pb collisions at $\sqrt{s_{NN}}=5.02~\mathrm{TeV}$. The shaded bands represent the uncertainty associated with the in-medium heavy-quark potentials. The experimental points are taken from the ALICE and CMS measurements~\cite{ALICE:2020wwx,CMS:2018zza}.}
\label{fig:raa_0_20}
\end{figure*}

Fig.~\ref{fig:raa_0_20} compares the bottomonium \(R_{AA}\) obtained with fluctuating and smooth hydrodynamic backgrounds in the 0--20\% and 30--50\% centrality intervals. The expected hierarchy, $R_{AA}(1S) > R_{AA}(2S) > R_{AA}(3S)$, is preserved in both backgrounds, reflecting the different in-medium stability of the three states. 
As illustrated in the figure, the discrepancy between the two background models remains relatively marginal in the 0-20\% centrality interval. 
For both centrality intervals, the effects of event-by-event medium fluctuations on the angle-integrated bottomonium $R_{\rm AA}$ are found to be marginal compared with the smooth reference calculation. The marginal discrepancy observed for $\Upsilon(1S)$ aligns with its tighter binding and reduced sensitivity to local temperature inhomogeneities. Similarly, investigations into the $\Upsilon(2S)$ and $\Upsilon(3S)$ states reveal only minor differences between the smooth and fluctuating medium scenarios. Therefore, as reflected in Fig.~\ref{fig:raa_0_20}, event-by-event fluctuations in the hot QCD medium do not lead to a quantitatively significant modification of the angle-integrated observable $R_{AA}$. 

The different \(p_{\rm T}\) dependence from Ref.~\cite{Song:2013tla} can be traced partly to the different treatment of the hard-production profile. Moreover, because the present calculation does not impose a sharp dissociation temperature, local hot spots do not remove a bottomonium state instantaneously; instead, they modify the accumulated wave-function evolution through the temperature-dependent complex potential. In the present work, the initial \(b\bar b\) positions are sampled from the same optical-Glauber binary-collision profile for both the fluctuating and smooth backgrounds. Therefore, the event-wise correlation between hard-production hot spots and medium hot spots is not included. This reduces the sensitivity of the ensemble-averaged \(R_{\rm AA}\) to local medium fluctuations across the entire \(p_{\rm T}\) range. For weakly bound excited states at low \(p_{\rm T}\), the suppression is already strong in the smooth background, leaving less room for additional visible reduction. At higher \(p_{\rm T}\), while individual trajectories through varying local temperature structures introduce a larger statistical variance, the continuous quantum evolution effectively averages out these spatial inhomogeneities, so that the central $R_{\rm AA}$ remains close to the smooth reference within the potential-induced uncertainty band.

\begin{figure*}[htbp]
\centering
\includegraphics[width=0.3\linewidth]{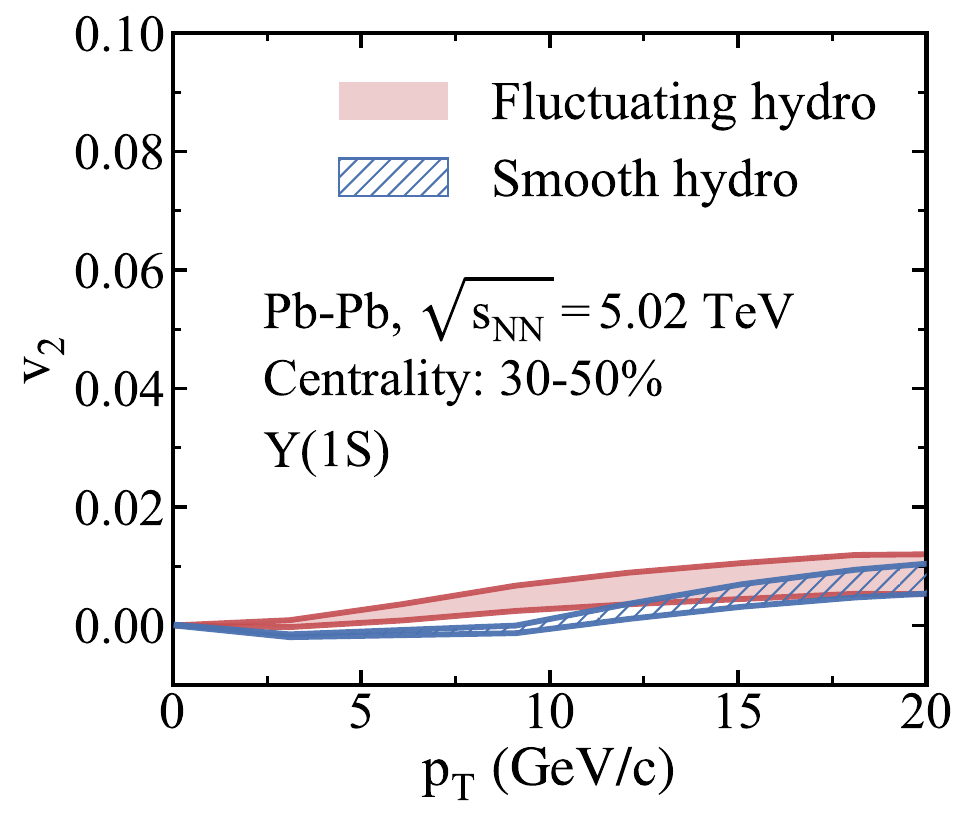}
\includegraphics[width=0.3\linewidth]{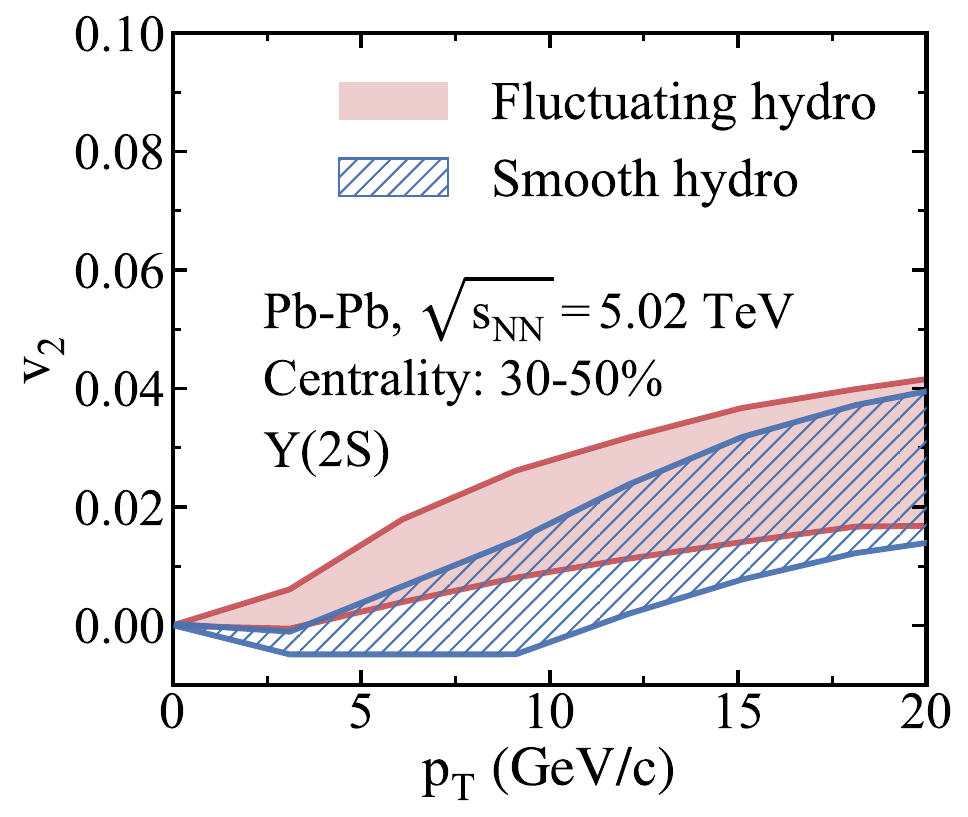}
\includegraphics[width=0.3\linewidth]{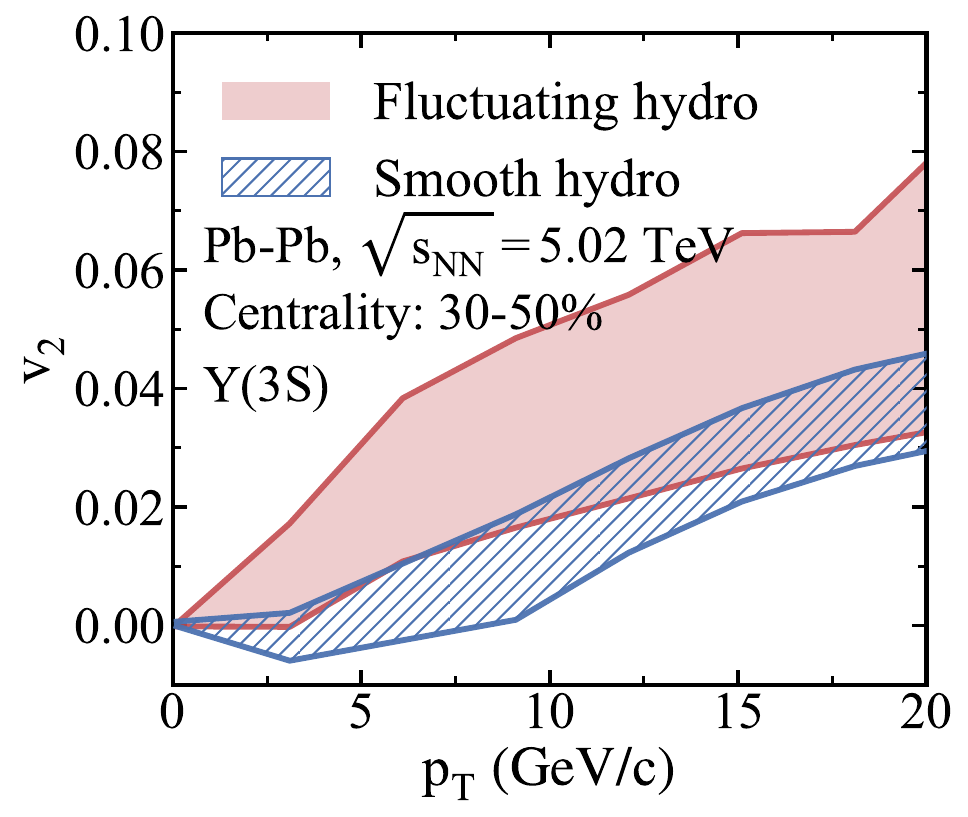}
\caption{Comparison of $v_2$ between fluctuating and smooth hydrodynamic backgrounds for $\Upsilon(1S)$, $\Upsilon(2S)$, and $\Upsilon(3S)$ in the 30--50\% centrality interval of Pb--Pb collisions at $\sqrt{s_{NN}}=5.02~\mathrm{TeV}$. The shaded bands are attributed to the uncertainty of the in-medium heavy-quark potentials.}
\label{fig:v2_30_50}
\end{figure*}

The corresponding comparison of \(v_2\) in the 30--50\% centrality interval is shown in Fig.~\ref{fig:v2_30_50}. The figure exhibits a clear state dependence: the elliptic flow remains very small for $\Upsilon(1S)$, becomes more visible for $\Upsilon(2S)$, and is largest for $\Upsilon(3S)$, showing the tendency,
$v_2(3S) \gtrsim v_2(2S) \gtrsim v_2(1S)$. 
In contrast to $R_{\rm AA}$, the elliptic flow shows a clear sensitivity to event-by-event fluctuations. The fluctuating background yields a systematically larger $v_2$ than the smooth background for all three states, and the difference becomes more pronounced for more weakly bound excited states. For $\Upsilon(1{\rm S})$ the enhancement is small and comparable to the potential-uncertainty band, whereas for $\Upsilon(2{\rm S})$ and $\Upsilon(3{\rm S})$ the fluctuating band lies systematically above the smooth band over the intermediate- and high-$p_{\rm T}$ region, with the separation for $\Upsilon(3{\rm S})$ exceeding the uncertainty propagated from the in-medium heavy-quark potential.

This behavior is consistent with the conventional relation between suppression and elliptic flow: the fluctuating and smooth backgrounds give a comparable angle-integrated $R_{\rm AA}$, while the fluctuating medium produces a larger $v_2$. The enhancement can be traced to the anisotropy that drives the bottomonium elliptic flow. In the smooth reference, $v_2$ is generated by the average, reaction-plane-like geometry of the optical-Glauber profile. In the fluctuating case, each event is analyzed with respect to its own second-order participant plane $\Psi_2$, whose eccentricity is enhanced by initial-state fluctuations; the resulting stronger and more irregular azimuthal temperature anisotropy along the trajectories leads to a larger differential suppression between the in-plane and out-of-plane directions. Because the more weakly bound excited states are more sensitive to the surrounding medium, this fluctuation-induced enhancement increases from $\Upsilon(1{\rm S})$ to $\Upsilon(3{\rm S})$.

\section{Summary}

In this work, we investigated the influence of event-by-event hydrodynamic fluctuations on bottomonium suppression and elliptic flow in Pb--Pb collisions at $\sqrt{s_{NN}}=5.02$ TeV. The internal quantum evolution of the bottomonium states was described by a time-dependent Schr\"odinger equation incorporating a temperature-dependent complex heavy-quark potential. The QGP background was simulated using the iEBE-VISHNU event-by-event viscous hydrodynamic framework. By comparing fluctuating and smooth hydrodynamic backgrounds for the $\Upsilon(1S)$, $\Upsilon(2S)$, and $\Upsilon(3S)$ states, we established a systematic picture of how hydrodynamic fluctuations modify the observables $R_{AA}$ and $v_2$.

Our results show that event-by-event fluctuations have only a marginal influence on the bottomonium $R_{\rm AA}$, reflecting the large masses and significant binding energies of the states and the fact that the angle-integrated yield is largely insensitive to local temperature inhomogeneities once the smooth background is constrained to the same charged-particle multiplicity. The elliptic flow $v_2$, by contrast, is systematically enhanced by the fluctuations, with the enhancement increasing for the more weakly bound excited states, because the participant-plane anisotropy of the fluctuating medium is larger than the average geometry of the smooth background. Bottomonium $v_2$ therefore retains a discernible sensitivity to event-by-event medium fluctuations, even though $R_{\rm AA}$ is only marginally affected. Higher-order bottomonium flow harmonics and two-particle correlation observables are expected to be more sensitive to event-by-event medium fluctuations than the angle-integrated \(R_{AA}\) and the geometry-dominated \(v_2\). A dedicated study of these observables requires higher statistics and will be pursued in future work.


\end{document}